\documentclass[twocolumn,preprintnumbers,amsmath,amsfonts,amssymb,notitlepage,superscriptaddress,showpacs,aps,prb,longbibliography]{revtex4-2}

\usepackage{color}

\usepackage{amsmath}

\usepackage{graphicx}

\usepackage{hyperref}

\def \be{\begin{equation}}
\def \ee{\end{equation}}
\def \ba{\begin{array}}
\def \ea{\end{array}}
\def \bea{\begin{eqnarray}}
\def \eea{\end{eqnarray}}

\newcommand{\khd}[1]

\date{\today}
\begin{document}
\title{Dicke materials as a resource for quantum squeezing}

\author{Vaibhav Sharma}
\email{vaibhavsharma@rice.edu}
\affiliation{Smalley-Curl Institute, Rice University, Houston, TX 77005, USA}
\affiliation{Department of Physics and Astronomy, Rice University, Houston, TX 77005, USA}

\author{Shung-An Koh}
\affiliation{Department of Materials Science and NanoEngineering, Rice University, Houston, TX 77005, USA}

\author{Jonathan Stepp}
\affiliation{Department of Physics and Astronomy, Rice University, Houston, TX 77005, USA}

\author{Dasom Kim}
\affiliation{Max Planck Institute for the Structure and Dynamics of Matter, Luruper Chaussee 149, 22761 Hamburg, Germany}

\author{Takumu Obata}
\affiliation{Interfaculty Graduate School of Innovative and Practical Studies, Yokohama National University, 79-8 Tokiwadai, Hodogaya-ku, Yokohama 240-8501, Japan}

\author{Yuki Saito}
\affiliation{Department of Physics, College of Engineering Science, Yokohama National University, 79-5 Tokiwadai, Hodogaya-ku, Yokohama 240-8501, Japan}

\author{Motoaki Bamba}
\affiliation{Department of Physics, College of Engineering Science, Yokohama National University, 79-5 Tokiwadai, Hodogaya-ku, Yokohama 240-8501, Japan}
\affiliation{Department of Physics, Graduate School of Engineering Science, Yokohama National University, 79-5 Tokiwadai, Hodogaya-ku, Yokohama 240-8501, Japan}
\affiliation{Institute for Multidisciplinary Sciences, Yokohama National University, 79-5 Tokiwadai, Hodogaya-ku, Yokohama 240-8501, Japan}

\author{Han Pu}
\affiliation{Smalley-Curl Institute, Rice University, Houston, TX 77005, USA}
\affiliation{Department of Physics and Astronomy, Rice University, Houston, TX 77005, USA}

\author{Hanyu Zhu}
\affiliation{Smalley-Curl Institute, Rice University, Houston, TX 77005, USA}
\affiliation{Department of Physics and Astronomy, Rice University, Houston, TX 77005, USA}
\affiliation{Department of Materials Science and NanoEngineering, Rice University, Houston, TX 77005, USA}
\affiliation{Department of Electrical and Computer Engineering, Rice University, Houston, TX 77005, USA}

\author{Junichiro Kono}
\affiliation{Smalley-Curl Institute, Rice University, Houston, TX 77005, USA}
\affiliation{Department of Physics and Astronomy, Rice University, Houston, TX 77005, USA}
\affiliation{Department of Materials Science and NanoEngineering, Rice University, Houston, TX 77005, USA}
\affiliation{Department of Electrical and Computer Engineering, Rice University, Houston, TX 77005, USA}

\author{Kaden R. A. Hazzard}
\affiliation{Smalley-Curl Institute, Rice University, Houston, TX 77005, USA}
\affiliation{Department of Physics and Astronomy, Rice University, Houston, TX 77005, USA}

\begin{abstract}
   We study magnetic materials whose low energy physics can be effectively described by a Dicke model, which we term \textit{Dicke materials}. We show how a Dicke model emerges in such materials due to a coexistence of fast-dispersing and slow-dispersing spins, which are strongly coupled. Analogous to the paradigmatic Dicke model describing light-matter interactions, these materials also exhibit signatures of a superradiant phase transition. The ground state near the superradiant phase transition is expected to be squeezed, making Dicke materials a resource for quantum metrology and witnessing entanglement in solid-state systems. However, as an entanglement measure, squeezing can be sensitive to perturbations that are otherwise irrelevant for usual correlation functions and order parameters. Motivated by the prospect of observing squeezing in such Dicke materials, we study the robustness of ground state squeezing under ubiquitous imperfections such as finite temperature, disorder, and local interactions. Using analytical and numerical techniques, we show that the squeezing obtained is perturbatively stable against these imperfections and quantitatively evaluate regimes promising for experimental observation. 
    
\end{abstract}

\maketitle

\section{Introduction}

Squeezed quantum states have emerged as powerful metrological tools that can enhance the measurement precision of physical quantities beyond the standard quantum limit~\cite{squeezereview}. Squeezed states have been used for such precision in gravitational wave detectors~\cite{ligo}, atomic clocks~\cite{clock1,clock2}, and atom interferometry~\cite{interferometry}. Spin squeezing additionally serves as an entanglement witness and measure~\cite{witness,witness1}. There is a vast body of work aimed at studying squeezed quantum states along with several experimental realizations in cold atoms, trapped ions and cavity QED systems~\cite{squeezereview, uscreview1,uscreview2,uscreview3}. Much work has focused on generating squeezed states using dynamical protocols of parametric driving~\cite{sqd1,sqd2,sqd3}, rotating atomic gases~\cite{rotating1,rotating2}, and Hamiltonian evolution~\cite{spinsq1,spinsq2,spinsq3,spinsq4,spinsq5,spinsq6,spinsq7,spinsq8,trappedion2,squeezedefn}. However, squeezed states can also be found in equilibrium, particularly in the ground states of special models. Ground state spin squeezing has been found in the symmetry breaking phase of the short-range XXZ models~\cite{qcpsq6} as well as near quantum critical points in higher-dimensional Ising models~\cite{qcpsq1,qcpsq2,qcpsq5} and Lipkin-Meshkov-Glick models~\cite{qcpsq3,qcpsq4}.

In addition to spin models, the ground state of the Dicke model -- a paradigmatic model of cavity QED with a single bosonic mode coupled to non-interacting spins -- near the normal-superradiant phase transition (SRPT) is squeezed~\cite{dickeanalytic,dickesqref1}. Exactly at the SRPT critical point, it has been shown that a two-mode photon-spin quadrature is perfectly squeezed~\cite{perfectsqueezing}. The closely related quantum Rabi model also hosts a  perfectly squeezed ground state at the critical point~\cite{rabi1}. The Dicke and the quantum Rabi models have been implemented in cavity QED experiments~\cite{coldatom1,coldatom2,coldatom3,coldatoms5,sccircuits} and trapped ions~\cite{trappedion,trappedion2,rabi2,rabi3}. 

Recently, the Dicke model has emerged as an effective description of the spectrum in certain orthoferrite materials~\cite{orthoferr,orthoferr2,orthoferr3,orthoferr4}. These are exemplars of what we term as \textit{Dicke  materials}. In such materials, one fast-dispersing mode plays the role of the photon-like bosonic degree of freedom (iron in the orthoferrites), while other magnetic degrees of freedom (erbium in the orthoferrites) act as the weakly-coupled spins. This interplay of fast and slow exchange interactions can simulate models of light-matter interaction without some limitations such as no-go theorems imposed by quantum electrodynamics~\cite{orthoferr}. This complements the strong ongoing effort to couple materials to real cavity photons, which can also realize Dicke models (albeit augmented with an $A^2$ term that significantly affects the physics)~\cite{uscreview1,uscreview2}. Together, these directions motivate the possibility of observing Dicke model quantum squeezing in solid-state systems and witnessing macroscopic entanglement in equilibrium for potential quantum metrology applications. 

In this context, it is important to consider equilibrium squeezing's robustness to unavoidable deviations from the ideal Dicke model that occur in solids, such as temperature, disorder, and local spin-spin interactions. While the consequences of such deviations have been considered for the equilibrium phase
diagrams in Dicke models~\cite{dickedisorder}, their effect on squeezing remains an open
question. Especially since squeezing is a sensitive observable and intimately connected to entanglement -- and thus can in principle behave quite differently from ordinary correlations measured in solids -- understanding the impact of these terms is crucial to observing quantum squeezing in solid-state systems. Previous works have studied the effect of disorder~\cite{disorder} and dissipation~\cite{dissipation} on spin squeezing in long-range spin models. In this work, we analyze how crucial deviations from the Dicke model affect the squeezed ground state near criticality. We find that quantum squeezing survives small and even experimentally realistic values of these perturbations. 

In Sec.~\ref{sec:dickematerial}, we describe the general anatomy of a Dicke material and characterize the conditions that lead to the emergence of a Dicke model. In Sec.~\ref{sec:model}, we study squeezing within the ideal Dicke model. This serves as a review of previously known results and helps us summarize key results in a language that we will use when we study Dicke models augmented by experimentally relevant deviations. We analytically derive the optimal squeezed quadrature as a function of model parameters at zero temperature in both the normal and superradiant phase. We also discuss squeezing in the presence of an additional $A^2$ term that is relevant for spin materials coupled to optical cavities where the familiar no-go theorem could apply.  

We then consider three unavoidable types of imperfections (temperature, disorder, and local spin-spin interactions) that are expected to be the most important physical effects. We study representative cases of each type. In Sec.~\ref{sec:temperature}, we show that squeezing decreases with temperature but survives up to a finite temperature. At fixed temperatures, it increases monotonically as the spin-boson coupling approaches the zero-temperature critical point, while it depends non-monotonically on detuning (difference of spin and boson excitation frequency).
In the latter case, squeezing is optimized at a finite distance away from the critical point at finite temperatures.

\begin{figure}
\includegraphics[width=\columnwidth]{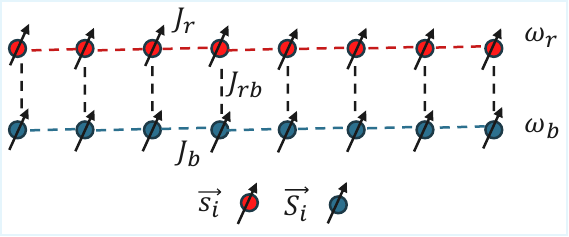}
    \caption{A schematic of a Dicke material. The two rungs host two spin degrees of freedom colored red and blue. The red (blue) spins have exchange interactions amongst themselves with strength $J_{\text{r}}$ ($J_{\text{b}}$) with excitation frequency $\omega_r$ ($\omega_b$). The red and blue spins have interspecies exchange interactions of strength $J_{\text{rb}}$. In a Dicke material, the $J_{\text{r}}$ is a dominantly large energy, providing a large velocity for excitations in the red leg analogous to the speed of light for photons in the Dicke model, while the $J_{\text{rb}}$ should also be significant enough to influence the physics of the blue spins.}
    \label{fig:dickematerial}
\end{figure}

In Sec.~\ref{sec:disorder}, we consider the effect of adding dilute disorder in the form of spins with disordered excitation frequencies and couplings. Disorder at some level is an inevitable feature of real experimental systems, especially in traditional condensed matter systems. 
We use perturbation theory to show that squeezing survives dilute disorder and the squeezed variance increases proportionally to the disorder fraction. We also use small system exact diagonalization numerics to go beyond the perturbative regime and confirm our finding. 

Finally, in Sec.~\ref{sec:interactions}, we consider the effect of local Ising-like spin interactions, which break the permutation invariance symmetry of spins in the ideal Dicke model. For weak Ising couplings, we analytically show that the spin excitations are magnon modes and the SRPT phase transition still occurs. The critical spin-boson coupling is simply renormalized by the Ising interactions and there is still perfect squeezing in the ground state. For strong Ising coupling, the ground state gets ferromagnetically ordered in the Ising coupling direction, destroying squeezing in the optimal quadrature. We use small system numerics to validate our results beyond the limit of small Ising couplings.

\section{ Dicke Materials}\label{sec:dickematerial}

Here we define the concept of Dicke materials and show how the Dicke model emerges as an effective description of these materials. A schematic of their general structure in one-dimension is shown in Fig.~\ref{fig:dickematerial}. The vertices on the top (bottom) rung shown in red (blue) are spin degrees of freedom that have ferromagnetic exchange interactions among themselves parametrized by $J_{\text{r}}$ ($J_{\text{b}}$). The two spin species also have exchange interactions with each other given by $J_{\text{rb}}$. For concreteness we assume each leg has SU$(2)$ symmetric interactions, but this is not a crucial ingredient. A simple example  Hamiltonian is given by $H_D = H_{\text{r}} + H_{\text{b}} + H_{\text{rb}}$ where
\begin{align}\label{eq:Hr}
    H_{\text{r}} &= -J_{\text{r}} \sum_i \vec{s}_i \cdot \vec{s}_{i+1} + \omega_r \sum_i s^z_i \\
\label{eq:Hb}
    H_{\text{b}} &= -J_{\text{b}} \sum_i \vec{S}_i \cdot\vec{S}_{i+1} + \omega_b \sum_i S^z_i \\
    H_{\text{rb}} & = \sum_{i,\alpha} J_{\text{rb}}^{\alpha} s^{\alpha}_i S^{\alpha}_{i}.
\end{align}
Here $\alpha \in \{x,y,z\}$ and $s^{\alpha}_i$ ($S^{\alpha}_i$) is the corresponding spin-1/2 operator on site $i$ on the top (bottom) rung of the lattice with excitation frequency along the $z$ axis given by  $\omega_r$ ($\omega_b$). The total number of sites per leg is $N$. 

The standard light-matter interaction Dicke model consists of a bosonic mode coupled to non-interacting spins. We will now show that when the couplings satisfy  certain conditions, $H_D$ reduces to such a Dicke model. 
In particular, as we will show, in the limit  $J_{\text{r}} \gg J_{\text{b}}$ and $J_{\text{b}} \ll \omega_b$, the blue spins can be described as non-interacting. In contrast, due to the red spins' strong exchange interactions, $J_{\text{r}}$, they can be described as fast dispersing magnon modes. Due to their fast dispersion, these modes are akin to the photonic mode in the standard Dicke model that can couple distant non-interacting spins. 

Formally, we can use  spin-wave theory and a Holstein-Primakoff transformation to write the low energy spectrum above the ferromagnetic ground state of the red spins in terms of bosonic magnon modes~\cite{lswt}. In the leading order of a Holstein-Primakoff transformation, $s^+_i \approx  a^\dagger_i$, $s^-_i \approx a_i$, and $s^z_i \approx 1/2$, where the $a_i$ ($a_i^\dagger$) are bosonic annihilation (creation) operators. Near the ferromagnetic ground state, this approximation is accurate. Substituting this in Eq.~\ref{eq:Hr} and doing a Fourier transform ($a_j = \sqrt{1/N} \sum_k e^{ikj} a_k$), the Hamiltonian is, up to additive constants, 
\begin{equation}
    H_{\text{r}} = \omega_k \sum_k a^\dagger_k a_k.
\end{equation}
Here $a^\dagger_k$, $a_k$ are creation and annihilation operator of the bosonic magnon modes for the red spins and $\omega_k = \omega_r + J_{\text{r}} (1-\cos k)$. The Hamiltonian for the blue spins is simply a Hamiltonian of non-interacting spins given by 
\begin{equation}
    H_{\text{r}} = \omega_b \sum_i S^z_i.
\end{equation}
For the interaction term, $H_{\text{rb}}$, to have interesting physics we require $J_{\text{rb}}$ to be comparable to $\omega_b$ for ``ultra-strong" spin-boson coupling~\cite{uscreview1,uscreview2}. Considering the interaction term along the $x$ and $y$ directions, we rewrite $s^x_i$ and $s^y_i$ in terms of the bosonic magnon modes defined above. The corresponding interaction term in $H_{\text{rb}}$ becomes
\begin{align}
    H_{\text{rb}} &= \frac{J^x_{\text{rb}}}{\sqrt{N}}\sum_m \sum_k (a_k e^{ikm} + a^\dagger_k e^{-ikm})S^x_m \nonumber \\
    &+\frac{i J^y_{\text{rb}}}{\sqrt{N}} \sum_m \sum_k (a^\dagger_k e^{-ikm}-a_k e^{ikm})S^y_m.
\end{align}
The interaction term denotes a linear collective coupling of the blue spins with bosonic magnon modes of momentum $k$. Note that the exchange coupling in the $z$-direction would give a non-linear term for the magnons due to the Holstein-Primakoff transformation. Near the ferromagnetic ground state of the red spins, this would be negligible compared to the exchange terms in the $x$ and $y$ directions.  

Put all together, the total Hamiltonian is
\begin{align}\label{eq:dickematmodel}
    H_D &= \sum_k \omega_k a^\dagger_k a_k + \omega_b\sum_m S^z_m \nonumber \\
    &+ \frac{J^x_{\text{rb}}}{\sqrt{N}} \sum_m \sum_k (a_k e^{ikm}+a^\dagger_k e^{-ikm})S^x_m \nonumber \\
    &+\frac{i J^y_{\text{rb}}}{\sqrt{N}} \sum_m \sum_k (a^\dagger_k e^{-ikm}-a_k e^{ikm})S^y_m
    .
\end{align}

Due to the ferromagnetic interactions among the red spins, the low energy mode is given by $k =0$. The different $k$ modes decouple, and for the $k=0$ mode, the effective Hamiltonian in Eq.~\ref{eq:dickematmodel} is a single mode Dicke model mathematically equivalent to the one used in the context of light-matter interactions. If one or more of these parameters are tunable by an external probe, one can study the Dicke model's paradigmatic superradiant phase transition (SRPT). In the language of spins, this transition will be equivalent to a magnetic ordering transition driven by the competition between the inter-species coupling $J_{\text{rb}}$ and $\{\omega_k,\omega_b\}$.

In summary, the Dicke model emerges as an effective description in a material that has two degrees of freedom with a separation of energy scales such that one of them disperses quickly while the other behaves like non-interacting spins.  We note that the calculations  immediately generalize to higher dimensions. The Dicke model also emerges as an effective model in cavity-coupled materials where the bosonic mode is the real photon mode of the cavity, and many of the situations described in this paper remain applicable in this scenario. However in these cavities, there is an $A^2$ term that can have non-trivial effects (see Sec.~\ref{sec:nogo}).

A  concrete examples of a Dicke material is Erbium Orthoferrite ($\text{ErFeO}_3$). This materials has been shown to exhibit an SRPT analogous to a Dicke model~\cite{orthoferr,orthoferr2,orthoferr3,orthoferr4}. In $\text{ErFeO}_3$, the Fe atomic spins strongly couple to each other, providing the fast dispersing magnon modes that play the role of the bosonic mode. The Er spins couple very weakly to each other, making them nearly independent spins. There is an anisotropic coupling between the Er and Fe spins, that can be interpreted as a collective coupling between the Fe magnon mode and the Er spins. The SRPT was observed by tuning the spin excitation frequency of the Er spins by a magnetic field~\cite{orthoferr}.

\section{Review of squeezing in the ideal Dicke model}\label{sec:model}

In this section, we review the physics of quantum squeezing in the ideal Dicke model. In Sec.~\ref{sec:diagonalization}, we diagonalize the Dicke model. In Sec.~\ref{sec:gssqueezing}, Sec.~\ref{sec:singlemodesqueezing}, and ~\ref{sec:sqawayfromcritpt}, we derive the quantum squeezing in the Dicke model. In Sec.~\ref{sec:spinsqueezing}, we review how bosonic squeezing is equivalent to the Kitagawa-Ueda spin squeezing criteria, and finally in Sec.~\ref{sec:nogo}, we discuss the effect of $A^2$ terms on squeezing in the Dicke model obtained from light-matter interactions. This section also introduces notation that will be used throughout the paper, especially when we consider the effect of deviations from the ideal Dicke model. 

The ideal Dicke model Hamiltonian describing a single bosonic mode coupled to $N$ non-interacting spin-1/2 particles is
\begin{equation}\label{eq:dickemodel}
    H/\hbar = \omega_0\sum_{i=1}^{N} S^i_z + \omega a^\dagger a + \frac{g}{\sqrt{N}}(a + a^\dagger)\sum_{i=1}^{N}(S^i_x).
\end{equation}
where $a,a^{\dagger}$ are the annihilation and creation operator of the bosonic mode. The spin-1/2 operators for the $i^{\text{th}}$ spin are $S^x_i,S^z_i$. The bosonic mode's excitation frequency is $\omega$, the spin excitation frequency along the $z$-axis is $\omega_0$, and the collective spin-boson coupling is $g$. 

We can diagonalize this Hamiltonian and find the energy spectrum by defining a collective spin-$N/2$ operator for the spins. We define the collective spin operators as
\begin{align}
    S_\alpha = \sum_{i=1}^{N} S^\alpha_i
\end{align}
where $\alpha \in \{x,y,z\}$. Due to the collective spin coupling, the Hamiltonian commutes with the total spin operator, $S^2 = S^2_x + S^2_y +S^2_z$. The eigenvalues of the $S^2$ operator are given by $s(s+1)$ where $s$ takes values in integer steps ranging from $0$ ($1/2$) to $N/2$ for even (odd) $N$. These eigensectors are called Dicke manifolds and they are not mixed by the Hamiltonian. The Hamiltonian also commutes with the parity operator, $\Pi = \exp{(i\pi P)}$ where $P = a^\dagger a + S_z + s$. The parity is thus conserved. 

In this section, we focus on the ground state in the totally symmetric sector where $s = N/2$. There are $N + 1$ eigenstates of the $S_z$ operator within this sector separated by equal energy spacings $\hbar\omega_0$. These energy spacings do not reflect the total energy and only correspond to the non-interacting spin Hamiltonian term. In the limit $N \to \infty$, this structure is reminiscent of a harmonic oscillator described by a bosonic creation and annihilation operators. For finite $N$,  this bosonization is formally carried out by the Holstein-Primakoff transformation where we define,
\begin{align}
    S_- = (\sqrt{N - b^\dagger b}) b \\
    S_+ = b^\dagger \sqrt{N - b^\dagger b} \\
    S_z = b^\dagger b - N/2,
\end{align}
where the $b$ and $b^\dagger$ are bosonic annihilation and creation operators. When $N \to \infty$ and we are near the $\langle S_z \rangle = -N/2$ state such that $\langle b^\dagger b\rangle \ll N$, we get, $S_- \sim \sqrt{N}b$ and $S_+ \sim \sqrt{N}b^\dagger$. Substituting this in Eq.~\ref{eq:dickemodel} and ignoring additive constants,  we get 
\begin{equation}\label{eq:bosonicdicke}
    H/\hbar = \omega_0 b^\dagger b + \omega a^\dagger a +  g(a + a^\dagger)(b^\dagger + b ) + \mathcal{O}(\frac{\langle b^\dagger b \rangle}{N}).
\end{equation}
This purely quadratic two-mode bosonic Hamiltonian, known as the Hopfield model, can be diagonalized into its normal modes. The parity operator becomes $\Pi = \exp{(i\pi (a^\dagger a + b^\dagger b))}$. 

\subsection{Diagonalizing into the normal modes}\label{sec:diagonalization}

The Hamiltonian in Eq.~\ref{eq:bosonicdicke} can be diagonalized using a Bogoliubov transformation. We will use an equivalent alternative approach by expressing the Hamiltonian in terms of position and momentum quadratures and then finding the normal modes. This will be helpful in later sections where we analyze the squeezing in these quadratures. 

We define canonically conjugate pairs of operators, $\{x,p_x\}$ and $\{y,p_y\}$ where
\begin{align}
    x = \frac{1}{\sqrt{2\omega}} (a+a^\dagger)\\
    p_x = i\sqrt{\frac{\omega}{2}}(a^\dagger - a)\\
     y = \frac{1}{\sqrt{2\omega_0}} (b+b^\dagger)\\
    p_y = i\sqrt{\frac{\omega_0}{2}}(b^\dagger - b).
\end{align}
In terms of these operators, Eq.~\ref{eq:bosonicdicke} becomes
\begin{equation}\label{eq:dickequad}
    H/\hbar = \frac{1}{2}(\omega^2 x^2 + p_x^2 + \omega_0^2 y^2 + p_y^2 + 4g\sqrt{\omega \omega_0}xy).
\end{equation}

We can now define canonically-conjugate pairs of normal mode operators, $\{q_{\pm},p_{\pm}\}$, such that
\begin{align}
    x = q_- \cos \gamma + q_+ \sin \gamma \\
    y = -q_- \sin \gamma + q_+ \cos \gamma \\
    p_x = p_- \cos \gamma + p_+ \sin \gamma \\
    p_y = -p_- \sin \gamma + p_+ \cos \gamma.
\end{align}
When $\tan 2\gamma = 4g\sqrt{\omega\omega_0}/(\omega_0^2-\omega^2)$~\cite{dickeanalytic}, the Hamiltonian decouples and becomes
\begin{equation}\label{eq:dickequad1}
    H/\hbar = \frac{1}{2}(\epsilon_-^2 q_-^2 + p_-^2 + \epsilon_+^2 q_+^2 + p_+^2)
\end{equation}
with
\begin{equation}\label{eigenenergy}
    \epsilon_{\pm}^2 = \frac{1}{2}\left( \omega^2 + \omega_0^2 \pm \sqrt{(\omega_0^2-\omega^2)^2+16g^2\omega\omega_0} \right).
\end{equation}

We can further define the new normal mode ladder operators $c_{\pm},c^\dagger_{\pm}$ where 
\begin{align}
    q_{\pm} = \frac{1}{\sqrt{2\epsilon_{\pm}}} (c_{\pm}+c_{\pm}^\dagger)\\
    p_{\pm} = i\sqrt{\frac{\epsilon_\pm}{2}}(c_{\pm}^\dagger - c_{\pm}).
\end{align}
In terms of these ladder operators, the Hamiltonian in Eq.~\ref{eq:dickequad1} becomes
\begin{equation}\label{eq:bosonicdicke1}
    H/\hbar = \epsilon_- c_-^\dagger c_- + \epsilon_+ c_+^\dagger c_+.
\end{equation}

The eigenstates of this Hamiltonian can be written in the number basis and represented as $|n_-,n_+\rangle$ where $n_j = c^\dagger_j c_j$ is the number of excitations in mode $j$ and thus $H/\hbar = n_- \epsilon_- + n_+ \epsilon_+$. The ground state corresponds to $n_-=n_+=0$.

\subsection{Ground state squeezing near the superradiant phase transition}\label{sec:gssqueezing}

A bosonic mode is squeezed when the variance of a quadrature falls below the value in a minimum uncertainty coherent state. Consider the Hamiltonian in Eq.~\ref{eq:bosonicdicke} and Eq,~\ref{eq:dickequad} with $g=0$. We then have two uncoupled oscillators with the ground state variances given by $\Delta x^2 (0) = 1/(2\omega)$, $\Delta p_x^2 (0) = \omega/2$, $\Delta y ^2 (0) = 1/(2\omega_0)$, and $\Delta p_y^2 (0) = \omega_0/2$. If the variance of a quadrature falls below the minimum of these ground state variances, we define that quadrature to be squeezed. 

In the Dicke model, one of the normal modes gets squeezed. This is most apparent near the normal to superradiant phase transition (SRPT) in the Dicke model that occurs at the critical coupling, $g=g_c$~\cite{dickeanalytic} where
\begin{equation}
    g_c = \frac{\sqrt{\omega\omega_0}}{2}.
\end{equation}

At this point, $\epsilon_- \to 0$ and $\epsilon_+ \to \sqrt{\omega^2 + \omega_0^2}$. In Eq.~\ref{eq:dickequad1}, this implies that the $q_-$ oscillator mode has a vanishing frequency. The ground state corresponds to the lowest eigenvalue of the $p_-^2$ operator, which is $0$ and the variance of this operator vanishes. This zero momentum state is infinitely extended in the $q_-$ coordinate while being perfectly squeezed in the $p_-$ coordinate as its variance is lower than the ground state variance of the uncoupled parent oscillators' quadratures.

We can see this more formally by calculating the variance of the $q_{\pm}$ and $p_{\pm}$ quadratures in the ground state. The ground state is given by the state $|0,0\rangle$ where $c_{\pm}|0,0\rangle= 0$. Using expressions of $q_{\pm},p_{\pm}$ in terms of ladder operators $c_{\pm}$, we find that the variance of the $q_{\pm}$ quadrature is $\Delta q_{\pm}^2 = \langle q_{\pm}^2 \rangle - \langle q_{\pm} \rangle^2 = 1/(2\epsilon_{\pm})$ and of the $p_{\pm}$ quadrature is $\Delta p_{\pm}^2 = \langle p_{\pm}^2 \rangle - \langle p_{\pm} \rangle^2 = \epsilon_{\pm}/2$. At the SRPT critical point, $\epsilon_- \to 0$, showing that the $p_-$ coordinate is perfectly squeezed with a vanishing variance. The $q_-$ quadrature is perfectly anti-squeezed with a variance that tends to infinity. In terms of the original bosonic modes, the squeezed quadrature is 
\begin{equation}\label{eq:twomode}
    p_- = i\sqrt{\frac{\omega}{2}} \cos \gamma  (a^\dagger - a) - i\sqrt{\frac{\omega_0}{2}} \sin \gamma  (b^\dagger - b).
\end{equation}

We can use the leading order Holstein-Primakoff transformation to write this quadrature in terms of the spin operators to get
\begin{equation}\label{eq:twomodespins}
    p_- = i\sqrt{\frac{\omega}{2}} \cos \gamma  (a^\dagger - a) - i\sqrt{\frac{\omega_0}{2N}} \sin \gamma  \sum_{j=1}^{N}(S^+_j - S^-_j).
\end{equation}
The relative weight on the bosonic and the spin modes is controlled by the angle $\gamma$ that depends on the model parameters. Perfect squeezing in this quadrature in the ground state at the critical point has also been found previously in Ref.~\cite{perfectsqueezing}.

We now define a new dimensionless parameter called the squeezing ratio, $\xi$, for a quadrature such that whenever $\xi < 1$ there is some quantum squeezing in that quadrature. Without loss of generality, we will consider $\omega < \omega_0$ throughout this paper unless specified otherwise. As a result, $\Delta p_x^2 (0) < \Delta p_y^2 (0)$, making $\Delta p_x^2 (0)$ the minimum ground state variance among the uncoupled oscillators. Thus, the squeezing ratio of the quadrature $p_-$ is defined as $\xi = \Delta p_-^2/\Delta p_x^2 (0)$. When $\xi < 1$, $p_-$ is squeezed. At the SRPT, $\xi \to 0$, indicating perfect squeezing. Intuitively, the finite coupling $g$ has created this new normal mode quadrature, $p_-$, whose variance is lower than the minimum ground state variance of the parent oscillator. This denotes squeezing in our case. We later show that the definition of bosonic squeezing of the Holstein-Primakoff boson in the Dicke model coincides with the usual definition of spin squeezing. 

\subsection{Single mode squeezing near the SRPT}\label{sec:singlemodesqueezing}

Near the SRPT critical point, not only is the two-mode quadrature $p_-$ squeezed, but so are the single mode quadratures, $p_x, p_y$ of the original bosonic modes. Unlike $p_-$ however, they are not perfectly squeezed. The variance of these single-mode quadratures in the ground state are
\begin{equation}
    \Delta p_x^2 = \langle p_x^2 \rangle - \langle p_x \rangle^2 = \frac{\epsilon_-}{2}\cos^2 \gamma + \frac{\epsilon_+}{2}\sin^2 \gamma 
\end{equation}

\begin{equation}
    \Delta p_y^2 = \langle p_y^2 \rangle - \langle p_y \rangle^2 = \frac{\epsilon_-}{2}\sin^2 \gamma + \frac{\epsilon_+}{2}\cos^2 \gamma. 
\end{equation}

At the critical point, $\epsilon_- = 0$, $\epsilon_+ = \sqrt{\omega^2 + \omega_0^2}$ and $\tan 2\gamma = 2\omega\omega_0/(\omega_0^2-\omega^2)$. Thus we get $\Delta p_x^2 = \omega^2/(2\sqrt{\omega^2+\omega_0^2})$ and $\Delta p_y^2 = \omega_0^2/(2\sqrt{\omega^2+\omega_0^2})$. In the ground state with zero coupling of the modes ($g=0$), the ground state variances would instead be $\omega/2, \omega_0/2$. We can see that the new variances are lowered by a factor of $\omega/\sqrt{\omega^2+\omega_0^2}$ and $\omega_0/\sqrt{\omega^2+\omega_0^2}$, leading to some squeezing in the $\{p_x,p_y\}$ quadratures. This agrees with the slight squeezing result in the $p_y$ quadrature found in Ref.~\cite{dickeanalytic}.  

Depending on the parameters of the Dicke model, the relative weight of the bosonic mode and the spins in the perfectly squeezed quadrature, $p_-$, can vary. For instance, when $\omega \gg \omega_0$, we get $\gamma \approx \pi/2$ and thus, $p_- \approx -i \sqrt{\omega_0/2} (b^\dagger - b)$. The perfectly squeezed quadrature comes almost entirely from the atomic spins. Conversely, the limit $\omega \ll \omega_0$ makes the bosonic mode quadrature the perfectly squeezed quadrature. The key factor that still causes squeezing in these extreme limits is the emergence of the new normal mode $q_-$ whose energy $\epsilon_-$ must be smaller than the smallest excitation frequency out of $\omega,\omega_0$. Near the phase transition, it always approaches zero regardless of the relative values of $\omega,\omega_0$. 

\subsection{Squeezing away from the SRPT critical point}\label{sec:sqawayfromcritpt}

While the squeezing is perfect at the SRPT critical point, there is squeezing even away from the critical point. As long as $\epsilon_- < \omega$ (true when $g \neq 0$), we have $\xi < 1$, signifying squeezing. 

\subsubsection{Within the normal phase}

In the normal phase, the ground state of the Dicke model is unique and has a fixed parity. It is a $+1$ eigenstate of the parity operator, $\Pi$, defined in Sec.~\ref{sec:model}. The ground state is the lowest energy state of the diagonalized Hamiltonian given in Eq.~\ref{eq:bosonicdicke1}.
For simplicity, we can consider the resonant case where $\omega = \omega_0$. Then the eigenenergy $\epsilon_-$ from Eq.~\ref{eigenenergy} becomes, $\epsilon_- = \omega\sqrt{1-2g/\omega}$. In the normal phase, $0<g<g_c$, we have $\xi = \sqrt{1-2g/\omega}< 1$, demonstrating squeezing. This can be seen by the blue curve in Fig.~\ref{fig:gsanalyticplot}. The squeezing is enhanced as we approach the critical point where $g/\omega = 1/2$. 

\subsubsection{Within the superradiant phase}\label{sec:sqsrphase}

When $g > g_c$, the system is in the superradiant phase. The ground state has a finite expectation value of the bosonic operators $a,b$ such that $\langle a \rangle = a_0$ and $\langle b \rangle = -b_0$. Here $a_0,b_0 > 0$ each serve as an order parameter of the superradiant phase transition. The ground state no longer has fixed parity and it is two-fold degenerate. The other degenerate ground state has the opposite sign of the order parameter such that, $\langle a \rangle = -a_0$ and $\langle b \rangle = b_0$. 

\begin{figure}
\includegraphics[width=\columnwidth]{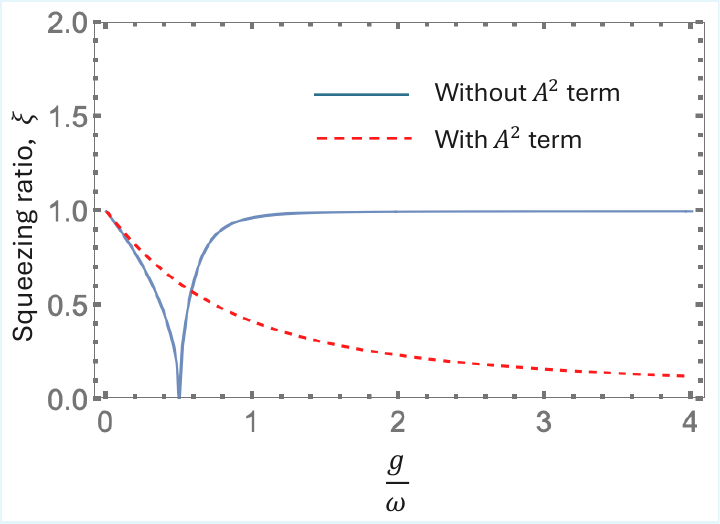}
    \caption{Squeezing ratio, $\xi$ as a function of $g/\omega$ at resonance ($\omega = \omega_0$) in the ground state of the Dicke model in Eq.~\ref{eq:bosonicdicke} (solid blue curve) and with the inclusion of the $ A^2$ term in the Dicke model in Eq.~\ref{eq:bosonicdickea2} (dashed red curve) with its numerical coefficient, $D = g^2/\omega$ . The squeezing ratio goes to zero (perfect squeezing) at $g/\omega=1/2$ only when there is no $A^2$ term.}
    \label{fig:gsanalyticplot}
\end{figure}

A valid bosonic Hamiltonian in this phase can be found by displacing the bosonic modes in Eq.~\ref{eq:bosonicdicke} by the order parameter values, $a_0,b_0$, which are determined to minimize the ground state energy. The resultant Hamiltonian in terms of the displaced bosonic modes can be further diagonalized in a manner analogous to what we saw in Sec.~\ref{sec:model}. This procedure and the new diagonalized Hamiltonian is derived in Ref.~\cite{dickeanalytic}. We only quote the final results here. The Hamiltonian in the superradiant phase is written as
\begin{equation}\label{eq:bosonicdicke2}
    H/\hbar = \tilde\epsilon_- e_-^\dagger e_- + \tilde\epsilon_+ e_+^\dagger e_+.
\end{equation}
Here $e_{\pm}$ are the creation and annihilation of the two oscillators defined around the superradiant ground state. The excitation energies, $\tilde \epsilon_\pm$ are
\begin{equation}
    \tilde\epsilon_\pm^2 = \frac{1}{2}\left( \omega^2 + \frac{g^4}{g_c^4}\omega_0^2 \pm \sqrt{\left(\left(\frac{g^4}{g_c^4}\omega_0^2-\omega^2\right)^2+4\omega^2\omega_0^2\right)}\right ).
\end{equation}
As we approach the critical point ($g \to g_c$), the excitation energy $\tilde \epsilon_-$ vanishes, signaling the SRPT. For simplicity, we again consider $\omega = \omega_0$. The squeezing ratio in the ground state is given by $\xi = \tilde \epsilon_-/\omega$. This is shown by the blue curve in Fig.~\ref{fig:gsanalyticplot} where we plot $\xi$ as a function of $g/\omega$, including both the normal phase ($g/\omega < 1/2$) and the superradiant phase ($g/\omega > 1/2$). Deep in the superradiant phase when $g \gg g_c$, $\tilde \epsilon_- \to \omega$ and thus $\xi \to 1$, similar to the normal phase when $g \to 0$. The squeezing is enhanced as we get closer to the critical point within the superradiant phase 

\subsection{Equivalence of spin squeezing and bosonic squeezing}\label{sec:spinsqueezing}

We show here that the bosonic squeezing of the collective atomic spins is equivalent to spin squeezing. For $N$ spins, the spin squeezing parameter $\xi$ is defined as~\cite{squeezedefn}
\begin{equation}\label{eq:spinsqueezing}
    \xi = \frac{4\Delta(S_{n\perp})^2}{N}.
\end{equation}
Here $ \Delta (S_{n_\perp})^2$ is the spin variance along a direction ($n_\perp$) perpendicular to the mean spin direction. There is spin squeezing when $\xi < 1$. In the normal phase before the critical point, the mean spin direction in the Dicke model is along the quantization direction, that is, the $z$-axis.
Consider $n_\perp = \hat y$ such that $S_{n_\perp} = i/2(S_+ - S_-)$ where $S_+$ and $S_-$ are collective spin raising and lowering operators respectively.  When $N \gg 1$, we get $S_+ \sim \sqrt{N} b^\dagger$ and $S_- \sim \sqrt{N} b$ by the Holstein-Primakoff transformation. Substituting this in Eq.~\ref{eq:spinsqueezing} and using the definition of the $p_y$ quadrature, we get
\begin{equation}
    \xi = \Delta\left(b^\dagger - b \right)^2 = \frac{2 \Delta p_y^2}{\omega_0}.
\end{equation}

\begin{figure}
\includegraphics[width=\columnwidth]{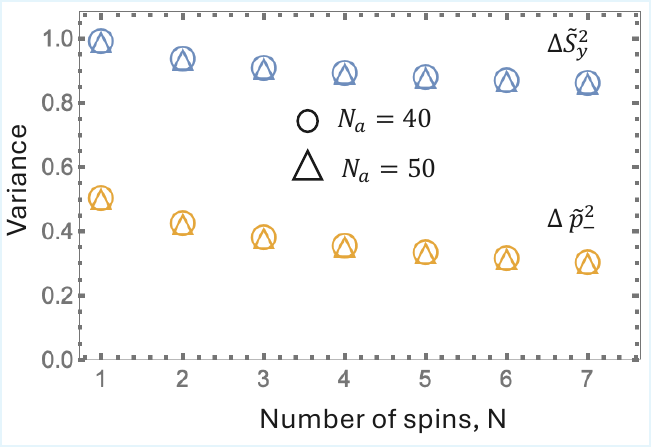}
    \caption{Ground state variance through exact diagonalization of operators $\tilde p_-$ (orange figures) and $\tilde S_y$ (blue figures) as a function of number of spins, $N$. Variance below 1 denotes squeezing and the squeezing gets better with increasing $N$. The circle and triangles show calculation with different boson truncation numbers, $N_a=40$ and $N_a=50$ respectively.}
    \label{fig:finitesizesq}
\end{figure}

The spin squeezing condition ($\xi < 1$) is equivalent to $\Delta p_y^2 < \omega_0/2$, which is the condition of bosonic squeezing.
Thus squeezing in the $p_y$ quadrature is equivalent to the traditional spin squeezing in the $S_y$ direction. Although this equivalence is strictly exact only when $N \to \infty$, we use exact diagonalization numerics to show that both the two-mode and the spin squeezing remarkably survive down to $N=2$ spins, although the exact values for small values of $N$ are different from the thermodynamic limit calculation. We numerically diagonalize the Dicke model Hamiltonian in Eq.~\ref{eq:dickemodel} with number of spins ranging from $N=1$ to $N=7$. We truncate the number of possible excitations of the bosonic field to $N_{\text{max}} = 40,50$. We consider the resonant case where $\omega = \omega_0$ and analyze the ground state at the thermodynamic limit critical point where $g = 0.5\omega$. 

We calculate the ground state variance of the operators $\tilde p_- = i/\sqrt{2} (a^\dagger - a) - i/\sqrt{2 N}(S_+ - S_-)$ and $\tilde S_y = i/\sqrt{N}(S_+ - S_-)$. Here $\tilde p_-$ is the finite size analog of the optimally squeezed bosonic quadrature, $p_-$, where one of the modes is the bosonic mode but the other mode is the exact spin operator as opposed to its Holstein-Primakoff boson counterpart. Similarly, $\tilde S_y$ is the exact spin operator analog of the bosonic quadrature $p_y$. Note that in the thermodynamic limit, $\tilde S_y \to p_y$ and $\tilde p_- \to p_-$.  There is squeezing when the variances of these operators are below 1. Fig.~\ref{fig:finitesizesq} shows the variances of the two operators as a function of the number of spins $N$ at the Dicke model critical point. We see that squeezing is present in both operators for all values of $N > 1$, and as $N$ becomes larger. As $N \to \infty$, our analytical results indicate that $\Delta \tilde p_-^2 \to 0$ and $\Delta \tilde S_y^2 \to 1/\sqrt{2}$. The scaling of the optimal squeezed quadrature for larger values of $N$ within the totally symmetric sector has also been shown in Ref.~\cite{perfectsqueezing}. 

\subsection{Dicke model squeezing and the no-go theorem}\label{sec:nogo}

When the bosonic mode that we considered in the Dicke model in Eq.~\ref{eq:dickemodel} is an electromagnetic field, we cannot neglect the squared electromagnetic potential term $A^2$. Inclusion of this term has been argued to prohibit the classical and quantum phase transitions in the Dicke model~\cite{nogo1,nogo2,nogo3,nogo4}, a result referred to as the no-go theorem. Adding the $A^2$ term in the Hamiltonian, the Dicke model from Eq.~\ref{eq:bosonicdicke} becomes
\begin{equation}\label{eq:bosonicdickea2}
    H/\hbar = \omega_0 b^\dagger b + \omega a^\dagger a +  g(a + a^\dagger)(b^\dagger + b ) + D(a+a^\dagger)^2
\end{equation}
in the large $N$ limit for the spins after the Holstein-Primakoff transformation. The term with coefficient $D$ is the $\hat{A}^2$ term. Diagonalizing this Hamiltonian as in Sec.~\ref{sec:diagonalization}, the eigenenergies are
\begin{equation}
    \epsilon_{\pm}^2 = \frac{1}{2}( \tilde\omega^2 + \omega_0^2 \pm \sqrt{(\omega_0^2-\tilde\omega^2)^2+16\tilde g^2\tilde\omega\omega_0} ).
\end{equation}
Here $\tilde\omega = \sqrt{\omega(\omega+4D)}$ and $\tilde g = g/(1+4D/\omega)^{1/4}$. When $0<D<g^2/\omega_0$, the SRPT occurs at a critical coupling $\tilde g_c$ given by $\tilde g_c = \sqrt{\tilde\omega\omega_0}/2$. At $\tilde g_c$, the excitation energy $\epsilon_-$ vanishes. However if the spins are considered as atomic dipoles coupled to a quantized electromagnetic field, the TRK sum rule gives $D = g^2/\omega_0$~\cite{nogo1}. In that case, the excitation energy $\epsilon_-$ never vanishes, signaling an absence of SRPT, i.e. there is no real value of $g$ for which the equation, $\tilde g = \sqrt{\tilde\omega\omega_0}/2$ is satisfied. 

Although there is no SRPT, the normal mode still superposes the spins and the electromagnetic field, leading to some squeezing. The squeezing ratio for this normal mode is given by $\xi = \epsilon_-/\omega$. The red curve in Fig.~\ref{fig:gsanalyticplot} shows $\xi$ as a function of $g/\omega$ at resonance ($\omega=\omega_0$). As the coupling $g$ increases, the squeezing gets enhanced and asymptotically approaches the perfectly squeezed case where $\xi = 0$. We need much higher values of coupling $g$ to get appreciable squeezing that is otherwise achieved near the SRPT critical point. 

We want to emphasize that this no-go theorem and the absence of the SRPT transition only applies when we consider spins as atomic dipoles coupled to an electromagnetic field, due to its reliance on the TRK sum rule. As we saw earlier, in Dicke materials, the role of the bosonic mode is played by a magnon mode and the spin-boson coupling is not an electromagnetic but a spin-spin coupling. There is an SRPT transition in that case~\cite{orthoferr,orthoferr2,orthoferr3,orthoferr4} and we expect enhanced squeezing near the critical point at finite coupling values. In systems such as trapped ions, the bosonic mode is a motional mode that couples to spins where again the no-go theorem is inapplicable. 

\section{Squeezing at finite temperature}\label{sec:temperature}

We now analyze the variance of the squeezed quadrature at finite temperatures. At a finite temperature $T$, the system density matrix is given by, $\rho = e^{-\beta H}/Z$ where $\beta = 1/(k_B T)$, $Z = \text{Tr} [e^{-\beta H}]$, and $H$ is the system Hamiltonian. This represents an incoherent mixture of the different eigenstates of the Hamiltonian in Eq.~\eqref{eq:bosonicdicke1} weighted by the Boltzamnn factor. The variance of the maximally squeezed quadrature, $p_-$, is
\begin{equation}\label{eq:trace}
    \Delta p_-^2 = \text{Tr}[\rho p_-^2] - \left(\text{Tr}[\rho p_-] \right)^2.
\end{equation}

\subsection{Normal phase}

We  first consider finite-temperature squeezing within the normal phase. From our earlier discussion, we know that the collective spin symmetry in the Dicke model divides the $N$ spins into disconnected sectors characterized by their total collective spin, $s$. One can find the ground state in each of these sectors, which in the normal phase is unique and has a fixed parity. 

Due to  $\omega_0$, the ground states within these sectors are not equal in energy. It is easiest to understand when the bosonic coupling $g=0$. The lowest energy state is the spin polarized state in the totally symmetric sector where the total spin is $s = N/2$ for $N$ spins. The ground states in all other sectors are higher in energy by multiples of $\omega_0$. 
In the normal phase, if the temperature $T$ is low such that $k_B T \ll \omega_0$, only excitations within the $s=N/2$ sector are relevant. Therefore, following the arguments in Sec.~\ref{sec:model}, the diagonalized Hamiltonian in Eq.~\eqref{eq:bosonicdicke} encodes the low energy physics near the ground state in this limit.

The system density matrix is obtained from the Gibbs ensemble of the two normal mode oscillators found in Eq.~\eqref{eq:bosonicdicke1} after diagonalization. We can then calculate the variance $\Delta p_-^2$ of the squeezed quadrature using Eq.~\ref{eq:trace} to evaluate the squeezing ratio, $\xi$. We find
\begin{equation}\label{eq:tempsq}
\begin{split}
    \xi &=  \frac{\sum_{n=0}^{\infty} \frac{\epsilon_-}{\omega}(2n+1)e^{-\beta n \epsilon_-}}{\sum_{n=0}^{\infty} e^{-\beta n \epsilon_-}} \\
    &= \frac{\epsilon_-}{\omega} \text{coth}\frac{\beta \epsilon_-}{2}.
\end{split}
\end{equation}

We can first look at some limiting behaviors to understand the temperature dependence. When temperature $T \to 0$  $(\beta \to \infty)$, one finds $\text{coth}(\beta\epsilon_-/2) \sim 1 + 2e^{-\beta\epsilon_-}$. Thus the squeezing ratio becomes, $\xi \sim \frac{\epsilon_-}{\omega} (1+2e^{-\beta\epsilon_-})$. The leading order  recovers the ground state squeezing ratio. When temperature $T \to \infty$ $(\beta \to 0)$, we have $\text{coth}(\beta\epsilon_-/2) \sim 2/(\beta\epsilon_-)$ and $\xi \sim 2k_B T/\omega$. In the high temperature limit, the squeezing ratio increases linearly with temperature. We  now consider the general finite temperature case described by Eq.~\eqref{eq:tempsq} in more detail, first as a function of $T$ and $g$ while fixing  $\omega/\omega_0$, and then as a function of $T$ and $\omega_0/\omega$ while fixing $g$. We show that there are finite regions in the temperature and model parameter space where the squeezing ratio, $\xi < 1$.

\subsubsection{Varying $g$ at fixed $\omega_0/\omega$}

\begin{figure}
\includegraphics{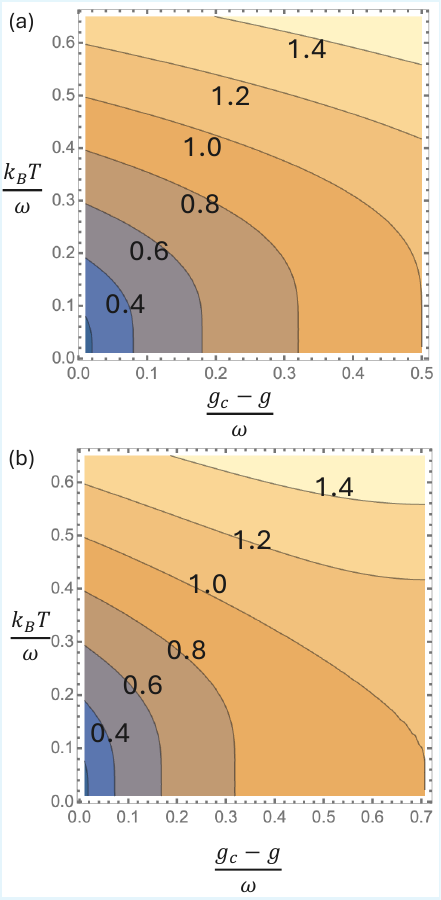}
    \caption{Contour plot of the squeezing ratio, $\xi$, in the normal phase ($g \leq g_c$) of the Dicke model as a function of temperature $T$ and $(g_c-g)$ when (a) $\omega_0=\omega$ and (b) $\omega_0 = 2\omega$. The contours are labeled by values of  $\xi$. Both cases are qualitatively similar and display a finite region in the parameter space where $\xi < 1$, indicating squeezing.}
    \label{fig:finitetempvaryg}
\end{figure}

We use  Eq.~\eqref{eq:tempsq} to plot the squeezing ratio as a function of $T$  and $g$ at fixed $\omega_0/\omega$. For simplicity, we consider $g < g_c = \sqrt{\omega\omega_0}/2$, always within the normal phase. Figs.~\ref{fig:finitetempvaryg}(a),(b) show the results of this calculation  for the resonant case ($\omega_0 = \omega$) and an off-resonant case ($\omega_0 = 2\omega$), respectively. In both cases, we find that there is no squeezing when $k_B T/\omega > 0.5$ as $\xi>1$ for those temperatures for all values of $g$. Squeezing is optimized as we either lower the temperature or move towards the SRPT critical point where $(g_c-g) \to 0$. 

Both the resonant and off-resonant cases are qualitatively similar. This is because the temperature scale is set by the lower excitation frequency out of $\omega$ and $\omega_0$. In the cases presented here, that frequency is $\omega$.
As we move away from the critical point, the contours bend downwards, reducing the threshold temperature beyond which $\xi > 1$.

\subsubsection{Varying $\omega_0/\omega$ at fixed $g$}

\begin{figure}
\includegraphics[width=\columnwidth]{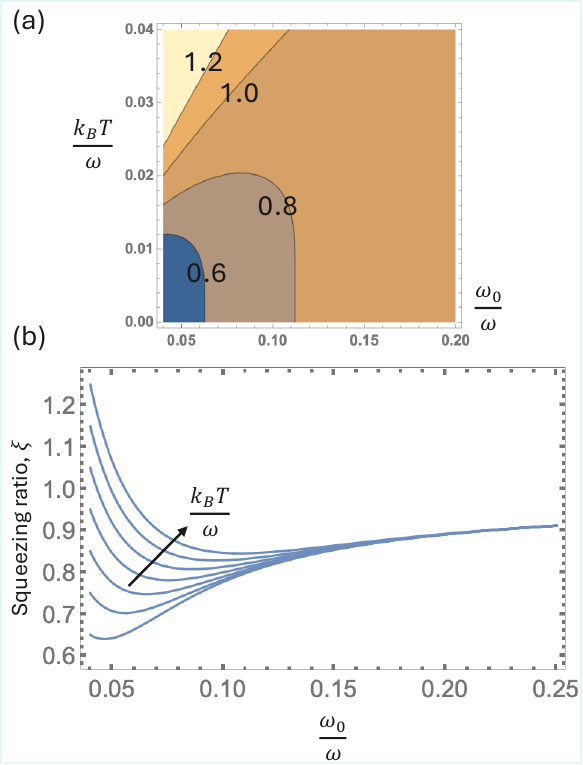}
    \caption{(a) Contour plot of the squeezing ratio, $\xi$, in the normal phase as a function of temperature $T$ and $\omega_0/\omega$ at a fixed value of $g=0.1\omega$. The contours are labeled by values of squeezing ratio, $\xi$. The $T=0$ SRPT critical point is at the bottom left corner where $\omega_0/\omega = 0.04$. (b) Squeezing ratio, $\xi$, as a function of $\omega_0/\omega$ at   temperatures ranging from $k_B T/\omega = 0.013$ to $0.025$ in steps of 0.002.
    }
    \label{fig:finitetempvaryomega}
\end{figure}

Now  we fix $g/\omega$ and vary $\omega_0/\omega$ and temperature, as shown in Fig.~\ref{fig:finitetempvaryomega}.  As an example, inspired by the realistic parameters in the Dicke material, $\text{ErFeO}_3$~\cite{orthoferr}, we consider  $g = 0.1\omega$, for which the $T=0$ phase transition occurs at $\omega_0/\omega = 0.04$. As  $\omega_0/\omega$ increases from this value, we move away from the SRPT critical point  deeper into the normal phase. Since $\omega_0 < \omega$, the squeezing ratio is defined by comparing the variance of the squeezed operator, $p_-$, with that of $p_y$ instead of $p_x$ in the earlier cases: $\xi =\Delta p_-^2/\Delta p_y^2$. At zero temperature, $\xi = \epsilon_-/\omega_0$.

Fig.~\ref{fig:finitetempvaryomega}(a) shows a contour plot of the squeezing ratio as a function of $T$ and $\omega_0/\omega$. The relevant temperature scale ($k_B T/\omega$) is much lower compared to the earlier cases because $\omega_0/\omega =0.04$ at the critical point, and the temperature scale is set by the lower excitation frequency, $\omega_0$ in this case. The behavior of $\xi$ versus $\omega_0/\omega$  at  fixed $T$ is  non-monotonic. Within a range of finite temperatures ($0.015 < k_B T/\omega < 0.02$), the minimum squeezing ratio is  at a finite distance away from the SRPT critical point.  Sufficiently close to the critical $\omega_0/\omega$, approaching it reduces the squeezing (increases $\xi$) due to growing thermal fluctuations.

This is also clearly seen in Fig.~\ref{fig:finitetempvaryomega}(b) where $\xi$ is plotted versus $\omega_0/\omega$ for various temperatures. The optimal squeezing occurs at a finite distance away from the critical point. As we lower the temperature, the minima reach lower values and occur closer to the critical point. Thus at $T>0$ in this case, the critical point is no longer guaranteed to be the optimal squeezing point. 

As a concrete illustration, we apply our results to $\text{ErFeO}_3$ modeled in Ref.~\cite{orthoferr3} with parameters $\omega = 2 \pi \times 0.896$ THz (43 K) and $g = 2 \pi \times 0.116$ THz. Based on our analysis above, squeezing is observable ($\chi < 1$) at temperatures up to $k_BT/\omega \approx 0.04$, corresponding to approximately 1.4 K — accessible with standard He-3 cryogenic equipment. At 100 mK in a dilution refrigerator, a squeezing depth of 9 dB ($\chi = 0.12$) is achievable. This provides a concrete experimental target for significant squeezing in a material that has already been characterized~\cite{orthoferr}.

\subsection{Superradiant phase}

From our discussion in Sec.~\ref{sec:sqsrphase}, we know that the ground state in the superradiant phase is two-fold degenerate and does not have a fixed parity. The low energy spectrum around each ground state is obtained by displacing the bosonic modes $a,b$ by their expectation values $a_0,b_0$ in the ground state. This is encoded by the quadratic Hamiltonian in Eq.~\eqref{eq:bosonicdicke2}.  The finite temperature behavior would then be analogous to our calculation within the normal phase if we restrict ourselves to this low energy subspace and fix $a_0$, $b_0$. However, the Dicke model is known to exhibit a finite temperature phase transition from the superradiant phase to the normal phase~\cite{classicalpt1,classicalpt2,classicalpt3,classicalpt4,classicalpt5} at $k_B T_c = \omega_0/ (2\tanh^{-1} (\omega\omega_0/4g^2))$~\cite{classicalpt4}. A finite temperature calculation within the low energy quadratic Hamiltonian derived in Sec.~\ref{sec:sqsrphase} does not exhibit this transition and thus it would not be valid for temperatures $T$ on the order of $T_c$ or higher. Moreover when $g \sim g_c$, $T_c \to 0$ implying that such a calculation would fail on the superradiant side for arbitrarily low temperatures as $g \to g_c$. We now discuss the reasons for these restrictions on the finite temperature calculation in the superradiant phase. 

The quadratic Hamiltonian in the superradiant phase in Eq.~\eqref{eq:bosonicdicke2} was derived by restricting to the $s=N/2$ total angular momentum sector for the spins (see Sec.~\ref{sec:model}). In this case, it has been shown that there is no thermal phase transition from superradiant to normal phase because there are not enough entropic fluctuations within a single total spin angular momentum sector to induce a thermal transition~\cite{classicalpt4,classicalpt5}. Only when we thermally average over all possible total spin sectors do we recover the thermal phase transition. 

We note that contrary to the normal phase where the different total spin sectors were gapped in energy by multiples of $\omega_0$, this is no longer true in the superradiant phase because the collective spin term $S_x$ is more dominant compared to $S_z$ in the Dicke hamiltonian. The spin eigenstates in $x$-basis are better descriptors of the low energy physics as opposed to the $z$-basis states. Similarly the bosonic field is also better described as a continuous variable coherent state as opposed to a discrete Fock state due to the dominant $(a+a^\dagger)$ term in the Dicke model Hamiltonian. Thus even temperatures lower than $\omega_0$ can mix different total spin sectors. An accurate thermal average for temperatures higher than $T_c$ would require the inclusion of all spin sectors. 

When $T \ll T_c$, the finite temperature squeezing results would be exactly analogous to the normal phase. In that case, the system density matrix is obtained from the Gibbs ensemble of the two normal mode oscillators (in Eq.~\ref{eq:bosonicdicke2}) describing the low energy subspace around the ground state. Our calculation breaks down when $T \sim T_c$. A detailed investigation of squeezing near this \textit{classical} phase transition, as well as across the full quantum critical regime, would be an interesting future topic for study.

\section{Squeezing with disordered spins}\label{sec:disorder}

We  now analyze the effect of  disorder on the ground state squeezing in the Dicke model. Generic disorder in the spins   would correspond to each spin having a random excitation frequency and coupling to the bosonic field. This destroys the collective spin symmetry, making it difficult in general to make analytical predictions about squeezing for a large number of spins as well as preventing efficient numerics. 

Our goal is to study the stability of squeezing to  disorder. In order to obtain an analytical handle on the problem, we restrict ourselves to dilute disorder. 
We consider a Dicke model where $N+m$ atomic spins interact with the bosonic field, out of which $N$ spins behave collectively while $m$ spins are defects with disordered excitation energies and couplings. This allows us to controllably vary the number of spins that break the collective spin approximation and perturbatively analyze  the system in the thermodynamic limit. Note that in the limit where $m \gg N$, we recover a fully disordered Dicke model.  Such a fully disordered Dicke model was studied in Ref.~\cite{dickedisorder} where it was found that the SRPT survives for sufficiently low disorder, but squeezing was not considered.

We  consider a Dicke model with dilute disorder, $m\ll N$, with the Hamiltonian $H =H_0 + H_d$ where
\begin{align}\label{eq:H0}
    H_0/\hbar &= \omega_0 \sum_{i=1}^{N} S^i_z + \omega a^\dagger a + \frac{g}{\sqrt{N+m}}(a + a^\dagger)\sum_{i=1}^{N}S^i_x. \\ 
    \label{eq:Hd}
    H_d/\hbar &= \sum_{i=N+1}^{N+m} \omega'_i S^i_z + \frac{(a + a^\dagger)}{\sqrt{N+m}}\sum_{i=N+1}^{N+m}g'_i S^i_x.
\end{align}
The $m$ disordered spins have independently randomly distributed excitation frequencies, $\omega_i'$, and couplings, $g'$ to the bosonic mode. Here we consider the case where the disordered spins' coupling direction ($x$) to the bosonic mode is perpendicular to their quantization axis ($z$). Our calculations will show that this choice has the maximum effect on the squeezed variance.


For the $N$ collective spins, we can follow the arguments of Sec.~\ref{sec:diagonalization} and diagonalize $H_0$ to be
\begin{equation}
    H_0/\hbar = \epsilon_- c_-^\dagger c_- + \epsilon_+ c_+^\dagger c_+.
\end{equation}
These are the same normal modes and  eigenenergies found in Sec.~\ref{sec:diagonalization} with the coupling  $g$ renormalized to $\bar g = g \sqrt{N/(N+m)}$. In the limit $m \ll N$, $\bar g \sim g$.  We write $H_d$ in terms of these normal modes. We know from Sec.~\ref{sec:diagonalization} that $x = \sqrt{1/2\omega}(a+a^\dagger) = q_- \cos \bar\gamma + q_+ \sin \bar\gamma$. Writing the normal mode quadratures $q_-,q_+$ in terms of the ladder operators $c_-,c_+$, we get
\begin{multline}
    H_d/\hbar = \sum_{i=N+1}^{N+m} \omega'_i S^i_z +
    \sqrt{\frac{\omega}{N+m}}  \\ 
    {}\times\left (\frac{\cos \bar\gamma}{\sqrt{\bar\epsilon_-}}(c_-+c_-^\dagger) + \frac{\sin \bar\gamma}{\sqrt{\bar\epsilon_+}}(c_++c_+^\dagger) \right)\sum_{i=N+1}^{N+m}g'_i S^i_x.
\end{multline}

Note that a bar over a variable indicates that the coupling $g$ has been renormalized to $\bar g$ in its calculation. At the SRPT transition for the $N$ collective spins, the coefficient of the coupling to the bosonic field in the disorder Hamiltonian becomes infinite since $\bar\epsilon_- \to 0$. This is because the disordered spins are coupling to the anti-squeezed mode. 
 If we are not exactly at the critical point, so that $\bar\epsilon_- \neq 0$, we can handle the effect of this term perturbatively. We will thus restrict our analysis to the normal phase where $\bar\epsilon_- \neq 0$. In the limit of $N \gg 1$ and $m \ll N$, if we have $g_i'\cos \bar\gamma\sqrt{\omega/((N+m)\bar\epsilon_-)} \ll \omega'_i$ for each disordered spin $i$, we can treat this coupling term in the disorder Hamiltonian perturbatively, since there are $m$ terms in the sum, and a $1/\sqrt{N+m}$ prefactor in the coupling. A perturbative approach is also adequate if $\omega \gg \omega_0$ where $|\cos \bar\gamma| \ll 1$, irrespective of $m/N$. The  total Hamiltonian can be written as $H = H_{\text{up}} + H_{\text{p}}$ where
\begin{align}
    H_{\text{up}}/\hbar &=  \bar\epsilon_- c_-^\dagger c_- + \bar\epsilon_+ c_+^\dagger c_+ + \sum_{i=N+1}^{N+m} \omega'_i S^i_z \\
    H_{\text{p}}/\hbar &= \alpha \left (\cos \bar\gamma(c_-+c_-^\dagger) + \sin \bar\gamma\sqrt{\frac{\bar\epsilon_- }{\bar\epsilon_+}}(c_+ + c_+^\dagger) \right) \nonumber
    \\ &\hspace{0.2in}{} \times \sum_{i=N+1}^{N+m}g'_i S^i_x.
\end{align}
Here $H_{\text{up}}$ is the unperturbed Hamiltonian,  $H_{\text{p}}$ is the perturbation, and  
\begin{equation}
    \alpha = \sqrt{\frac{\omega}{(N+m)\bar\epsilon_-}}. 
\end{equation}
The parameter $\alpha$ serves as a small parameter that lets us perturbatively expand our new ground state. The ground state for the unperturbed Hamiltonian, $|\psi_0\rangle$ is  
\begin{equation}
    |\psi_0\rangle = |0,0\rangle \otimes |s\rangle
\end{equation}
where $|0,0\rangle$ is the ground state of the $c_-,c_+$ normal modes, and $|s\rangle$ is a product state of the $m$ disordered spins aligned along the $z$-axis. If all the disordered excitation frequencies $\omega'$ are large and positive, $|s\rangle$ would be all spins aligned along the negative $z$-axis. If there are some negative values of $\omega'$, the corresponding defect spins would point up but the overall unperturbed state remains a product state. For now, let's consider all the $\omega'$ values to be positive.
The unperturbed ground state energy of this state is $E_0/\hbar = (\bar\epsilon_- + \bar\epsilon_+)/2 - \sum_{i=N+1}^{N+m}\omega'_n/2 $. Letting  $E_k$ be the energy of the $|k\rangle$ eigenstate of $H_{\text{up}}$, first order perturbation theory gives the ground state wavefunction
\begin{align}
    |\psi'\rangle &= |\psi_0\rangle + \sum_{k}\frac{\langle k | H_{\text{p}} | \psi_0 \rangle |k\rangle}{E_0 - E_k} \nonumber \\ 
    & = |\psi_0\rangle - \alpha \cos \bar\gamma|1,0\rangle \otimes \sum_{N+1}^{N+m}\frac{ g'_i S^i_x }{\bar\epsilon_- + \omega'_i}|s\rangle \nonumber \\
    &\hspace{0.2in}{}- \alpha \sin \bar\gamma \sqrt{\frac{\bar\epsilon_-}{\bar\epsilon_+}}|0,1\rangle \otimes \sum_{N+1}^{N+m}\frac{ g'_i S^i_x }{\bar\epsilon_+ + \omega'_i}|s\rangle.
\end{align}

\subsection{Squeezed quadrature including the disordered spins}

Recall that in the absence of disorder, the squeezed quadrature is simply given by the operator $p_- = i \sqrt{\bar\epsilon_-/2}(c^\dagger_--c_-)$. It is a two-mode quadrature containing the bosonic field operator and the collective atomic spin operators explicitly written in Eq.~\ref{eq:twomode}. In an actual system, some of the spins are disordered but one usually will not have experimental access specifically to the disordered spins. Thus, we calculate the squeezing when our two-mode quadrature includes these disordered spins. In terms of the bosonic field and spin operators, this quadrature $p_d$ is
\begin{equation}
    p_{d} = i\sqrt{\frac{\omega}{2}}\cos \bar\gamma (a^\dagger - a) -i\sqrt{\frac{\omega_0}{2(N+m)}}\sin \bar\gamma \sum_{n=1}^{N+m}(S^n_+ - S^n_-).
    \label{eq:disorderquad}
\end{equation}

Without disordered spins ($m=0$), we recover, $p_d = p_-$, as written in Eq.~\ref{eq:twomodespins}. We can separate out the $N$ collective spins to write $p_d$ as the sum of the original squeezed operator $p_-$ and a purely disordered spin operator. The operator $p_d$ then becomes,
\begin{equation}
    p_d = i\sqrt{\frac{\bar\epsilon_-}{2}}(c^\dagger_--c_-) - \sqrt{\frac{\omega_0}{2(N+m)}}\sin \bar\gamma \sum_{n=N+1}^{N+m} S^n_y
\end{equation}
where $S^n_y = i(S^n_+ - S^n_-)$. The squeezing ratio $\xi$ is given by $\xi = \Delta p_d^2/(\omega/2)$. By calculating the variance of the quadrature $p_d$ in the  perturbed ground state $|\psi'\rangle$, we find that  
\begin{align}
    \xi &= \frac{\bar\epsilon_-}{\omega} +\sin^2 \bar\gamma \frac{\omega_0}{\omega}\frac{m}{N+m} \nonumber\\
    &\hspace{0.1in}{}-\frac{\alpha \cos \bar\gamma}{\omega}\sqrt{\frac{\bar\epsilon_-\omega_0}{(N+m)}} \sum_{i=N+1}^{N+m} \frac{2\langle s| S^z_i|s \rangle g'_i}{(\bar\epsilon_- + \omega'_i)} \label{eq:disordersqratio}
\end{align}
to first order in $\alpha$.

We see that the squeezing ratio gets additive corrections that depend on the  distribution of the $\omega_i'$ and $g_i'$ as well as the disorder fraction, $m/(N+m)$. The first correction term is the direct contribution to the variance from the pinned disordered spins, while the last term arises from the boson-spin coupling. Note that this last term is non-zero only when the disordered spins couple to the bosonic field along the $x$ axis, maximizing the effect on the squeezing ratio.

The most important feature is that the disorder corrections scale as $O(m/(N+m))$, so are small  when the disorder fraction is small. In particular, the squeezing survives sufficiently close to the SRPT critical point. The second order perturbation theory contribution to the squeezing ratio scales as $m^2/(N+m)^2$ due to the summation of $m^2$ terms of order $\alpha^2$ with a $1/(N+m)$ prefactor coming from squaring the squeezed quadrature $p_d$. Thus, our first order perturbation theory treatment is valid.


When $\alpha g'_i/\omega'_i \sim 1$, we are no longer in the perturbative limit. This happens when a finite fraction of the disordered spins have small excitation frequencies, $\omega_i'$, or sufficiently large coupling, $g'_i$, to the bosonic field. The gap to the unperturbed excited states is of the order of $\omega' + k \bar\epsilon_-$ where $k$ counts the number of excitations of the $c_-$ mode. 
In such a scenario, when we are close to the critical point where $\bar\epsilon_-/\omega$ is small,  the disordered perturbation Hamiltonian  resonantly mixes large excitations of the $c_-$ mode and the disordered spin states in the ground state. There is significant entanglement generated between the disordered spins and the $c_-$ mode, and our perturbative truncation to leading order in $\alpha$ breaks down. The variance of the squeezed quadrature $p_d$ could potentially increase  to destroy any squeezing. For strong enough perturbations,  we would also expect there to be no squeezing in the original quadrature $p_-$ that includes only the $N$ ordered spins.  This has been shown by a monogamy argument where significant entanglement of the bosonic field with disordered spins reduces the squeezing caused by entanglement between the bosonic field and the $N$ ordered spins~\cite{monogamy}. In the next section, we address the non-perturbative physics.

\subsection{Exact numerics away from the perturbative disorder limit}

\begin{figure}
\includegraphics[width=\columnwidth]{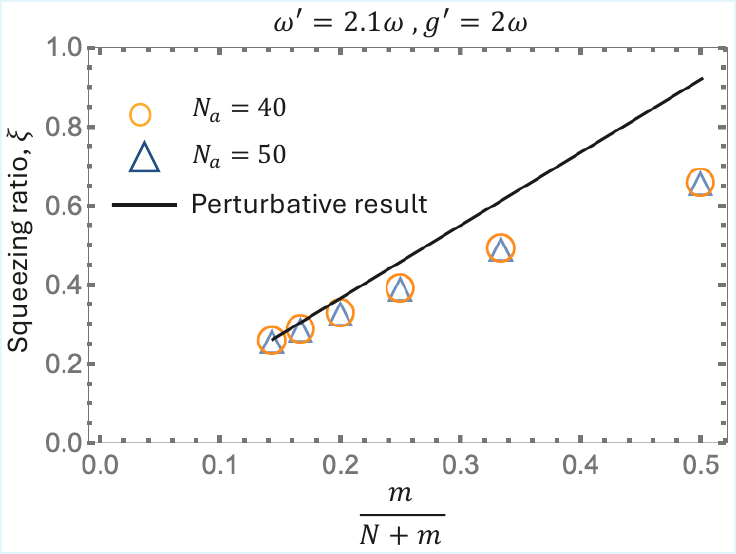}
    \caption{Ground state squeezing ratio, $\xi$, in the quadrature given in Eq.~\ref{eq:disorderquad} as a function of disordered spin fraction ($m/(N+m)$) with $m=1$ in the non-perturbative limit where $g'=2\omega$, $\omega' = 2.1\omega$ at the thermodynamic limit $T=0$ Dicke critical point ($\omega=\omega_0$ and $g = 0.5\omega$). The solid line plots the perturbative expression in Eq.~\ref{eq:disordersqratio}. The data points correspond to exact diagonalization numerics with boson truncation numbers, $N_a=40$ and $N_a = 50$ showing convergence in truncation.}
    \label{fig:disordersq}
\end{figure}

In order to understand the physics beyond the small $m/N$ or small $g'_i/\omega'_i$ limits,  we numerically calculate the squeezing in  small systems. We  exactly diagonalize  the  disordered Hamiltonian given by Eqs.~\ref{eq:H0},\ref{eq:Hd} where we fix to a single disordered spin, $m=1$. We vary the total number of collective spins from $N=1$ to $N=6$ and find the ground state on resonance ($\omega=\omega_0$) when the collective spin coupling corresponds to the thermodynamic critical point ($g = 0.5\omega$). In our numerics, we truncate the boson occupation to $N_{\text{max}} = 40,50$. 

Fig.~\ref{fig:disordersq} shows the squeezing ratio $\xi$ as a function of disorder fraction ($m/(N+m)$) in the non-perturbative limit where $g'/\omega' \sim 1$. We find that the general behaviour follows the perturbative disorder case, namely that the squeezing improves with lower disordered spin fraction. Our finite size numerics lends further credence to our understanding that squeezing survives finite disorder even beyond the simple perturbative limits we considered earlier and that squeezing is improved as the disorder becomes more dilute. 

 \section{Squeezing with local spin interactions: Dicke-Ising Model}\label{sec:interactions}

Finally we consider the effect of local spin interactions among the atomic spins on ground state squeezing. These local interactions are ubiquitous in materials and break the collective spin symmetry in the Dicke model, greatly complicating its theoretical treatment. To get an analytical understanding of their effect, we consider the case of nearest-neighbor ferromagnetic Ising spin interactions between the atomic spins. Our results are expected to be general for other forms of ferromagnetic local interaction, as long as they are weak. We consider a one-dimensional chain  with periodic boundary conditions for simplicity. The  Hamiltonian can be written as $H = H_c + H_M + H_I$ where
\begin{align}\label{eq:dickeising1}
    H_a &= \sum_k \omega_k a^\dagger_k a_k. \\
\label{eq:dickeising2}
    H_M &= \omega_0\sum_{n=1}^{N} S^n_z + 4J\sum_{n=1}^{N} S^n_x S^{n+1}_x. \\
\label{eq:hinteraction}
    H_I &= \frac{g}{\sqrt{N}} \sum_k \sum_n \left(a_k e^{ink} + a^\dagger_k e^{-ink} \right)S^n_x.
\end{align}

Here we consider multiple momentum modes of the bosonic field characterized by their excitation frequencies, $\omega_k$, in $H_a$. The matter Hamiltonian for the atomic spins is $H_M$ where $\omega_0$ is the spin excitation frequency along the $z$-axis and there are nearest-neighbor Ising interactions along the $x$-axis with periodic boundary conditions such that $S^{N+1}_x = S^1_x$.
The last part, $H_I$, is the interaction Hamiltonian that couples the different bosonic momentum modes collectively to the atomic spins. The Dicke-Ising model's ground state phase diagram has been studied in Refs.~\cite{dickeising1,dickeising2,dickeising3,dickeising4}. The Dicke-Ising model is exactly solvable using a mean field approach when the spins' collective coupling to the bosonic mode is along the same direction as the quantization axis ($z$)~\cite{dickeising4}. We will only consider the case where these two directions are orthogonal, as is typical in Dicke models. 

When $J=0$, we get a multimode Dicke model. Restricting to the $k=0$ mode leads to the standard Dicke model where the Hamiltonian separates into decoupled sectors defined by the total angular momentum, $s$, of the collective atomic spins. However, finite-$k$  modes break this collective symmetry due to their spatial variation, leading to non-zero matrix elements between different total angular momentum sectors. Similarly, when $J \neq 0$, the local spin-spin interaction terms break the conservation of total angular momentum. Therefore, we express the atomic spin excitations in terms of the $k$-modes.

Consider the lowest energy polarized spin state given by $|s\rangle = \otimes_{n=1}^{N}|\downarrow\rangle_n$. A magnon spin excitation of momentum $k$ can be created by an operator of the form $\sum_{n=1}^{N} e^{ink}S^n_+$. For a given bosonic mode $k$, the interaction term in Eq.~\ref{eq:hinteraction} creates or destroys a magnon spin excitation of momentum $k$. When $J = 0$, a magnon excitation of momentum $k$ costs an energy $\hbar\omega_0$ due to flipping an additional spin. This energy cost is independent of $k$, reflecting a large degeneracy in momentum.  However, when $J \neq 0$, this degeneracy in momentum is also broken.  In the next section, we  diagonalize the atomic spin part of the Hamiltonian in terms of magnon modes and then analyze how these can couple to the bosonic field to give ground state squeezing.

\subsection{Diagonalizing the matter Hamiltonian, $H_M$}

The matter Hamiltonian, $H_M$, is simply the 1D Ising model that can be exactly diagonalized by a Jordan-Wigner transformation into fermions~\cite{dickeising}. There is a phase transition at $\eta \equiv J/\omega_0 = 1$ where the spins order along the $x$-axis. To calculate the ground state squeezing in the Dicke-Ising model, we find it more convenient to describe $H_M$ in terms of bosonic magnon modes. We consider the $\eta \ll 1$ limit where the system can be described by magnon-like excitations on top of the all-down ground state.

We  perform a Holstein-Primakoff transformation~\cite{dickeising} for each atomic spin-1/2, $S^n_+ = b_n^\dagger \sqrt{1-b^\dagger_n b_n}$ and $S^n_- = \sqrt{1-b^\dagger_n b_n}\,  b_n $. This transformation is non-linear and gives a non-quadratic Hamiltonian, but
when $\eta \ll 1$, the low energy states have only small numbers of spin excitations, and thus $\langle S^n_z \rangle + 1/2 = \langle b^\dagger_n b_n \rangle \ll 1$. In this limit, we can truncate the Holstein-Primakoff transformation to  leading order; this can be verified by considering perturbation theory in the dropped terms noting their magnitude is higher order in $\eta$.  We note that the resulting magnon spectrum of the Ising model spectrum only matches the exact Ising model spectrum to linear order in $\eta$~\cite{dickeising}. As $\eta$ increases, one would need to go to higher orders to match the exact spectrum.


Truncating to the leading order in Holstein-Primakoff transformation such that $S^n_+ \sim b_n^\dagger$
and $S^n_- \sim  b_n$, we get
\begin{equation}\label{eq:ising}
    H_M = \omega_0 \sum_n b_n^\dagger b_n + J\sum_n (b^\dagger_n+b_n)(b^\dagger_{n+1}+b_{n+1}).
\end{equation}
We Fourier transform, 
\begin{equation}\label{eq:fourier}
    b_n = \frac{1}{\sqrt{N}}\sum_k e^{ink} b_k, 
\end{equation}
and  diagonalize the Hamiltonian into magnon modes by the Bogoliubov transformation
\begin{equation}\label{eq:bogol}
    d_k = \alpha_k b_k + \beta_k b^\dagger_{-k}.
\end{equation}
The  matter Hamiltonian becomes
\begin{equation}
     H_M = \sum_k E_k d^\dagger d_k
\end{equation}
where, up to linear order in $\eta$, one finds $E_k = \omega_0 (1+2\eta \cos k)$, $\alpha_k = 1$, and $\beta_k = \eta \cos k$~\cite{dickeising}. When $\eta = 0$, all magnon modes are degenerate with excitation frequency $\omega_0$ as expected for collective non-interacting spins, while a finite value of $\eta$ gives a finite magnon dispersion. For ferromagnetic Ising interactions ($J < 0$), the $k=0$ magnon mode is  lowest in energy while for antiferromagnetic interactions ($J > 0$), the lowest energy mode is $k=\pi$.




\subsection{Diagonalizing the full Dicke-Ising Hamiltonian}

We can now rewrite the Dicke-Ising Hamiltonian's interaction term, $H_I$, in terms of the magnon modes found above. The leading order Holstein-Primakoff transformation  $2S^n_x \sim b^\dagger_n + b_n$,  Fourier transform in Eq.~\ref{eq:fourier}, and the Bogoliubov transform in Eq.~\ref{eq:bogol} substituted into $H_I$ (Eq.~\ref{eq:hinteraction}) give
\begin{equation}
    H_I = g \sum_k (1+\eta \cos k)(a_{k}+a^\dagger_{-k})(d_k^\dagger + d_{-k}).
\end{equation}
The entire Hamiltonian is now quadratic with  momentum-dependent parameters coupling the matter magnonic modes with the bosonic modes, 
\begin{align}
    H &= \sum_k \omega_k a^\dagger_k a_k + \sum_k E_k d^\dagger_k d_k  \nonumber \\
    &\hspace{0.1in}{}+ g\sum_k (1+\eta\cos k)(a_{k}+a^\dagger_{-k})(d_k^\dagger + d_{-k}). \label{eq:dickeisingquad}
\end{align}
This can be diagonalized analogously to the Hamiltonian in Eq.~\ref{eq:bosonicdicke} in Sec.~\ref{sec:diagonalization}. For each mode $k$, we define position and momentum quadratures where $x_k = \sqrt{1/(2\omega_k)} (a_k + a_{-k}^\dagger)$, $p_{x,k} = i\sqrt{\omega_k/2}( a^\dagger_{-k}-a_k)$, $y_k = \sqrt{1/(2E_k)}(d^\dagger_k + d_{-k})$, and $p_{y,k} = i\sqrt{E_k/2}(d_k^\dagger-d_{-k})$. In terms of these quadratures, the Dicke-Ising Hamiltonian in Eq.~\ref{eq:dickeisingquad} becomes
\begin{align}\label{eq:dickeisingnormal}
    H &= \sum_{k>0} [( \omega_k^2 x_k x_{-k} + p_{x,k}p_{x,-k} + E_k^2 y_k y_{-k} + p_{y,k}p_{y,-k} \nonumber
    \\ 
    &\hspace{0.4in} {}-2\tilde g_k \sqrt{\omega_k E_k} (x_k y_{-k}+x_{-k}y_k)]
\end{align}
where $\tilde g_k = g(1+\eta\cos k)$. 
We  define normal mode coordinates $q_{\pm,k},p_{\pm,k}$ for each mode $k$ such that $x_k = q_{-,k} \cos \gamma_k + q_{+,k} \sin \gamma_k$, $p_{x,k} = p_{-,k} \cos \gamma_k + p_{+,k} \sin \gamma_k$, $y_k = -q_{-,k} \sin \gamma_k + q_{+,k} \cos \gamma_k $ and $p_{y,k} = -p_{-,k} \sin \gamma_k + p_{+,k} \cos \gamma_k$. We use the symmetry of the Hamiltonian in Eq.~\ref{eq:dickeisingnormal} when $k \to -k$ to impose that $\gamma_k = \gamma_{-k}$. 
Choosing  $\tan 2\gamma_k = 4\tilde g_k \sqrt{\omega_k E_k}/(\omega_k^2-E_k^2)$, the Hamiltonian becomes
\begin{equation}
\begin{split}
    H = \sum_{k>0} \big(\epsilon_{-,k}^2 q_{-,k}q_{-,-k} + p_{-,k}p_{-,-k} \\
   {} + \epsilon_{+,k}^2 q_{+,k}q_{+,-k}+p_{+,k}p_{+,-k} \big)
\end{split}
\end{equation}
where
\begin{equation}
    \epsilon_{\pm,k}^2 = \frac{1}{2}\left(\omega_k^2 + E_k^2 \pm \sqrt{(\omega_k^2-E_k^2)^2 + 16\tilde g_k^2 \omega_k E_k}\right).
\end{equation}

As we saw earlier in the original Dicke model, we can define ladder operators $c_{\pm,k}$ where $q_{\pm,k} = \sqrt{1/(2\epsilon_{\pm,k})}(c_{\pm,k}^\dagger + c_{\pm,-k})$ and $p_{\pm,k} = i\sqrt{\epsilon_{\pm,k}/2}(c_{\pm,k}^\dagger - c_{\pm,-k})$, in terms of which the Hamiltonian  is   given by
\begin{equation}
    H = \sum_{k>0} (\epsilon_{-,k}c^\dagger_{-,k}c_{-,k}+ \epsilon_{+,k}c^\dagger_{+,k}c_{+,k}).
\end{equation}

\subsection{Squeezing in the Dicke-Ising model}

We have now expressed the Dicke-Ising model entirely analogously to the original Dicke model but now summed over different momentum modes. Analogously, for each momentum mode $k$, one finds a separate SRPT when $\epsilon_{-,k} \to 0$. This occurs when $\tilde g_k = \tilde g_c(k) = \sqrt{\omega_k E_k}/2$. The first to occur will govern the true phase transition that occurs when the modes are weakly coupled. The corresponding normal mode, $c_{-,k}$, has the maximally squeezed quadrature   $p_{-,k} = i\sqrt{\epsilon_{-,k}/2}(c^\dagger_{-,k}-c_{-,-k})$.  The squeezing ratio in the case where $\omega_k < E_k$ is, $\xi = \epsilon_{-,k}/\omega_k$, with perfect squeezing at the SRPT transition (the case $\omega_k > E_K$ can be treated similarly). In terms of the parent bosonic modes, the squeezed quadrature is given by 
 a linear combination of the bosonic mode and the matter magnon mode
\begin{equation}
    p_{-,k} = i\sqrt{\frac{\omega_k}{2}} \cos \gamma_k  (a^\dagger_{-k} - a_k) - i\sqrt{\frac{E_k}{2}} \sin \gamma_k  (d^\dagger_k - d_{-k}).
\end{equation}  
Using Eq.~\ref{eq:fourier} and Eq.~\ref{eq:bogol},  the squeezed quadrature in  $b_n,b^\dagger_n$ to leading order is 
\begin{multline}
     p_{-,k} = i\sqrt{\frac{\omega_k}{2}} \cos \gamma_k  (a^\dagger_{-k} - a_k)\\
     - i\sqrt{\frac{E_k}{2N}} \sin \gamma_k (1-\eta \cos k)\sum_{n=1}^{N} e^{ink} (b^\dagger_n - b_n).
\end{multline}
and in terms of the original spin operators to leading order is
\begin{multline}
     p_{-,k} = i\sqrt{\frac{\omega_k}{2}} \cos \gamma_k  (a^\dagger_{-k} - a_k)\\
     - \sqrt{\frac{E_k}{2N}} \sin \gamma_k (1-\eta \cos k)\sum_{n=1}^{N} e^{ink} S^y_n.
     \label{eq:dickeisingsqquad}
\end{multline}

The squeezed quadrature rotates as a function of mode $k$ to linear order in $\eta$. Using the expression derived for $\tilde g_c(k)$ earlier, the critical coupling for mode $k$ up to leading order in $\eta$ is given by
\begin{equation}
    g_c(k) = \frac{\sqrt{\omega_k \omega_0 }}{2} + O(\eta^2).
\end{equation}
There is no shift in the critical coupling at linear order in $\eta$. If $\omega_k$ is a monotonically increasing function, the $k=0$ mode has the lowest critical coupling and is the most squeezed first as the coupling $g$ increases.

For larger Ising couplings $J$, we would need to consider further higher order terms in the Holstein-Primakoff transformation. Such terms would give interactions among the different magnon modes, and the squeezing is no longer simple to handle analytically.

\begin{figure}
\includegraphics[width=\columnwidth]{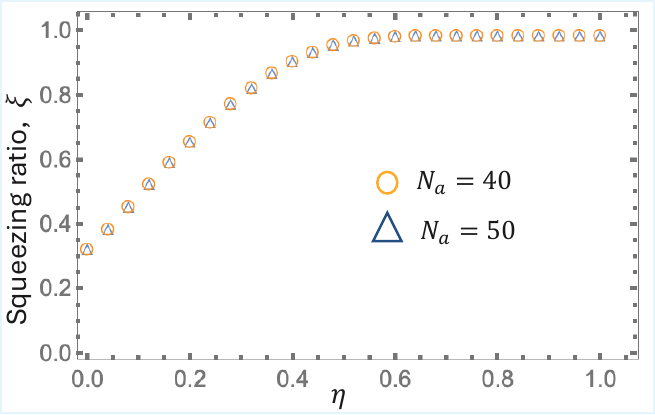}
    \caption{Squeezing ratio, $\xi$, for the quadrature in Eq.~\ref{eq:dickeisingquad} for $k=0$ as a function of Ising coupling, $\eta=J/\omega_0$, for $N=6$ spins at the $T=0$ thermodynamic-limit Dicke critical point  ($\omega_{k=0}=\omega_0$ and $g = 0.5\omega$). The data points correspond to exact diagonalization numerics with boson truncation numbers, $N_a = 40$ and $N_a = 50$ showing convergence in truncation.}
    \label{fig:dickeisingsq}
\end{figure}

We can explore such regimes through  small system exact diagonalization. We consider a system of $N=6$ spins and find the ground state of the Dicke-Ising model  in Eqs.~\ref{eq:dickeising1},\ref{eq:dickeising2},\ref{eq:hinteraction},  truncating the boson occupation to $N_{\text{max}} = 40, 50$. We calculate the variance of the operator $p_{-,k}$ given in Eq.~\ref{eq:dickeisingsqquad} for $k=0$, at the parameters for the $T=0$ thermodynamic-limit critical point, $g_c(k)$ for $k=0$ when $\omega_{k=0} = \omega_0$.



Fig.~\ref{fig:dickeisingsq} shows $\xi$ for this quadrature as a function of the Ising coupling $\eta$. We find that squeezing survives for small Ising interactions, and that  $\xi$ monotonically increases with $\eta$. When $\eta \sim 1$, the spins  align along the $x$-axis and are no longer coupled resonantly to the bosonc mode. Thus the squeezing disappears, and  $\xi$ saturates at $1$.

\section{Summary and Outlook}

We introduced the notion of \textit{Dicke materials} inspired by the recent experiment studying rare-earth orthoferrites~\cite{orthoferr}. In these materials, there are both fast dispersing and slow dispersing degrees of freedom. The faster ones are analogous to photons while the slower ones are analogous to localized spins. As a result, these materials are able to simulate ultra-strong light-matter interaction models such as the Dicke model in a real material without the requirement of actual strong light-matter coupling.   

The Dicke model hosts a superradiant phase transition with the ground state at the critical point showing perfect squeezing. This motivates the prospect of observing quantum squeezing in solid-state systems through Dicke materials. We studied squeezing in models relevant to such materials. In particular, we have studied squeezing including a number of important pertubations to the pure Dicke model: temperature, disorder, and local interactions. 

We first focused on the ground state of the ideal Dicke model and analytically derived the optimal two-mode squeezed quadrature as a function of the Dicke model parameters at zero temperature. We reproduce earlier findings that the two-mode squeezing is perfect at the super-radiant phase transition (SRPT) critical point. We also find  single mode squeezing in the bosonic field and the purely spin quadrature. 

We then consider the effect of some common experimental imperfections such as finite temperature, disorder, and local spin interactions on the squeezed ground state within the normal phase close to criticality. We show that squeezing survives up to a finite temperature scale that depends on the Dicke model parameters. At fixed temperature and $\omega/\omega_0$ the squeezing increases as we approach the critical coupling $g_c$. In contrast, when we fix $g$ and vary $\omega_0/\omega$, the optimal squeezing at a fixed finite temperature can occur away from the SRPT critical point. 

For disorder, we considered the effect of adding disordered spins that couple to the bosonic mode. In the perturbative limit where the disorder is dilute, we analytically showed that squeezing survives and the squeezed variance grows proportionally to the fraction of disordered spins. We used small system numerics to go beyond the perturbative regime and verified that squeezing persists for dilute or weak disorder. 

Finally, we derived the effect of local Ising spin-spin interactions that break the collective nature of spins in the original Dicke model. For weak interactions, we show that the spins can be described as magnons that couple to the bosonic mode.
The two-mode squeezing is retained in the regime. We corroborated our results with small system numerics to show that there is ground state squeezing in the presence of small Ising interactions. For simplicity, we have focused on the normal side of the transition ($g<g_c$) in all our calculations. These can be extended to the ordered superradiant phase by considering excitations around the two degenerate ground states. 

We have shown that quantum squeezing is not extremely fragile when the ideal Dicke model is perturbed and that it is present even for very small system sizes away from the thermodynamic limit. This makes it amenable to observation in Dicke materials~\cite{orthoferr,orthoferr2,orthoferr3,orthoferr4} as well as in quantum simulators~\cite{coldatom1,coldatom2,coldatom3,coldatoms5,sccircuits,trappedion,trappedion2,rabi2,rabi3}. In the context of solid-state systems, signatures of quantum squeezing and entanglement can be witnessed by experiments~\cite{condexprev} that measure magnetic susceptibility~\cite{condexp1}, spin-noise spectroscopy~\cite{condexp3}, and quantum Fisher information through inelastic neutron scattering~\cite{paschen,condexp2}. As a specific and experimentally relevant example, $\text{ErFeO}_3$ is predicted to exhibit ground-state squeezing observable up to 1.4 K, with a squeezing depth of 9 dB ($\chi = 0.12$) achievable at 100 mK in a dilution refrigerator.

This identification of a class of materials with the Dicke model opens up new avenues of research. On the one hand, characteristic phenomena of Dicke physics -- for example, the SRPT and entanglement and squeezing near it -- can be studied in real materials. On the other, the generalizations naturally presented by these materials combine local interactions with long-range effective interactions through the fast magnon mode, potentially enriching the physics well beyond the simple Dicke model and presenting new opportunities for theory to understand their interplay~\cite{uscreview3}. An interesting area of future study is to explore how broadly these  (rather non-stringent)  requirements can be realized and fruitfully applied in materials.
\\
\section{Acknowledgements}

V.S. thanks Sohail Dasgupta for insightful discussions and acknowledges support from the J. Evans Attwell-Welch fellowship by the Rice
Smalley-Curl Institute. K.H. acknowledges support from  the National Science Foundation DGE-2346014 and the W. M. Keck Foundation (Grant No. 995764). K.H.'s work was performed in part at the Aspen Center for Physics, which is supported by the National Science Foundation grant PHY-2210452. H.P. acknowledges support from the NSF (grant no. PHY-2513089) and the
Welch Foundation (Grant No. C-1669). M.B. acknowledges support from the Research Foundation for Opto-Science and Technology and from the Japan Society for the Promotion of Science (JSPS) Grant Number JPJSJRP20221202 and KAKENHI Grant Numbers JP24K21526, JP25K00012, JP25K01691, and JP25K01694.

\bibliography{main.bib}

\end{document}